\documentclass[journal,twoside,web]{ieeecolor}
\usepackage{generic}
\usepackage{cite}
\usepackage{amsmath,amssymb,amsfonts}
\usepackage{algorithmic}
\usepackage{graphicx}
\usepackage{subcaption}
\usepackage{textcomp}

\def\BibTeX{{\rm B\kern-.05em{\sc i\kern-.025em b}\kern-.08em
		T\kern-.1667em\lower.7ex\hbox{E}\kern-.125emX}}
\begin{document}
%\title{ Breaking the Si-FinFET scaling limit including Source-to-Drain Tunneling using Negative Capacitance}
\title{Physics and modeling of multi-domain FeFET with domain wall induced Negative Capacitance}
%Addressing Source to Drain Tunneling in Extremely Scaled Si-FinFET using Negative Capacitance  
\author{Nilesh Pandey, \IEEEmembership{Graduate Student Member, IEEE}, and Yogesh Singh Chauhan, \IEEEmembership{Fellow, IEEE}
\thanks{Nilesh Pandey, and  Yogesh Singh Chauhan are with Department of Electrical Engineering, Indian Institute of Technology - Kanpur,  Kanpur, 208016, India. e-mail: pandeyn@iitk.ac.in; chauhan@iitk.ac.in.}
\thanks{This work was supported
	in part by the Swarna Jayanti Fellowship under Grant DST/SJF/ETA02/2017-18 and in part by the Department of Science and Technology
	through the FIST Scheme under Grant SR/FST/ETII-072/2016.}
% <-this % stops a space
}

\maketitle 

\begin{abstract}
In this paper, we present the dynamics and modeling of multi-domains in the ferroelectric FET (FeFET). Due to the periodic texture of domains, the electrostatics of the FeFET exhibit an oscillatory conduction band profile. To capture such oscillations, we solve coupled 2-D Poisson's equation with the net ferroelectric energy density (gradient energy + free energy + depolarization energy) equation. Multi-domain dynamics are captured by minimizing the
net ferroelectric energy leading to a thermodynamically stable state. Furthermore, we show that the motion of domain walls originates local bound charge density in the ferroelectric region, which induces the negative capacitance (NC) effect. The strength of domain wall-induced NC is determined by the gradient energy of the ferroelectric material. FeFET exhibits variability in the
drain current with domain period due to the inherent NC effect.
Additionally, the impact of domain wall transition (soft$\rightleftharpoons$hard) on the device's electrostatic/transport is also analyzed. The model also accurately captures both nucleations of a new domain and the motion of the domain wall. Furthermore, the model is thoroughly validated against experimental results and phase-field simulations.
\end{abstract}
\begin{IEEEkeywords}
Multi-Domain, Domain dynamics, Domain wall induced negative capacitance, Ferroelectric energy dynamics, Green's function.
\end{IEEEkeywords}
\section{Introduction}
\label{sec:introduction}
\IEEEPARstart{F}{erroelectricity} discovery in the Hafnium oxide-based material led to the enormous research interest in the CMOS compatible ferroelectric FET (FeFET), which possesses the profound possibility to utilize it as an efficient non-volatile memory (NVM) \cite{Boscke}-\cite{Florent}. Ferroelectric layer in the FeFET exhibits multi-domain texture, which leads to an energy minima (thermodynamically stable) state  \cite{Bratkovsky}-\cite{yanchuk}. Numerous studies based on the phase-field (numerical) simulations have demonstrated that the density of domains in the FE region is determined by the interplay between various ferroelectric energy components\cite{Park}-\cite{Zubko}. Furthermore, the presence of negative capacitance (NC) effect via domain wall motion is observed in the ferroelectric material \cite{Bratkovsky_2}-\cite{Saha_4}. 
The concept of negative capacitance is extensively studied in FETs to break the fundamental limit of sub-threshold slope \cite{Salahuddin}-\cite{h}. Nevertheless, the physics of domain formation in the ferroelectric material is crystalline. But, the modeling (analytical/compact) and physics of domain dynamics in a FeFET remains elusive.

\begin{figure}[!t]
	\centering 
		\centering
		\includegraphics[width=0.5\textwidth]{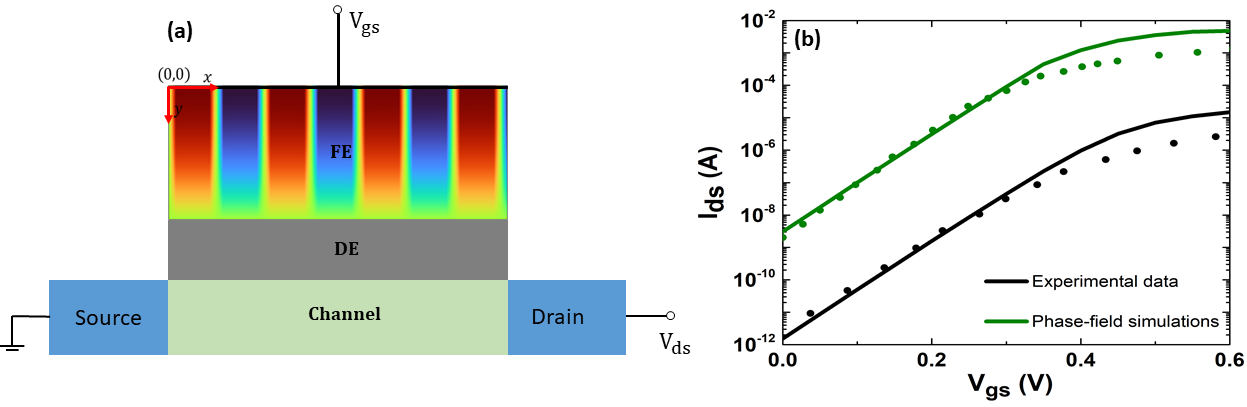}
	\caption{(a) Schematic of double gate FeFET (the half structure is shown). (b) Validations of the developed model with the phase-field simulations \cite{Saha_4} and experimental data \cite{Krivokapic}. The metal work-function and ferroelectric material parameters are tuned to calibrate the model. Default ferroelectric material parameters are taken from \cite{Saha_4}.     %Default parameters: $t_{si}$ =5 nm, $t_{ox}$ = 0.5nm, $\epsilon_{ox}$ = 3.9$\epsilon_0$, gate metal work-function= 4.4eV, $V_{gs}$ = 0 V, $V_{ds}$ = 0.6 V, undoped body, S/D doping = $10^{20}cm^{-3}$ and T=300 K.}% metal gate work function is 4.4} 
}\label{fig:str_1}\vspace{-2mm}
\end{figure}

It is well established that the interaction between various energy components determines the state and density of domains in a ferroelectric material \cite{Bratkovsky}-\cite{Rabe},\cite{Park}-\cite{Saha_4}. Therefore, we follow the same approach to capture both nucleation and motion of the domain wall. In this paper, we present an explicit analytical model of multi-domain FeFET. The 2-D Poisson's equation is solved with lateral and vertical gradients in the domain polarization. Subsequently, obtained electric field profile is used to calculate the ferroelectric layer's gradient and depolarization energy density. Net ferroelectric energy (free energy + gradient energy + depolarization energy) is minimized to model the dynamics of multi-domains.
Furthermore, the model can capture the secondary effects such as negative capacitance induced by domain wall motion and domain wall transition from soft $\rightleftharpoons$ hard type caused by ferroelectric thickness scaling. The negative capacitance effect is a vital function of domain wall width, which causes variability in the drain current characteristics. Additionally, the experimental results and phase-field simulations thoroughly validate the developed model and algorithm. Fig. \ref{fig:str_1}(a) shows the schematic of an double-gate FeFET with the coordinate axis used in this paper.
 \label{sec:introduction }
 
\begin{figure*}[!t]
	\centering 
	\includegraphics[width=1\textwidth]{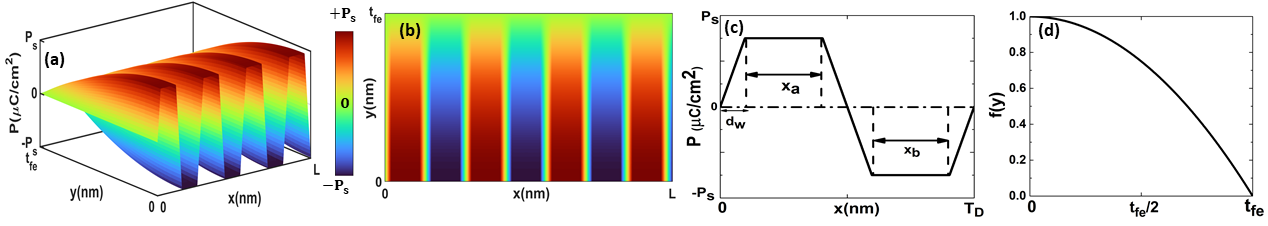}
	\caption{(a) 3-D polarization distribution in the FE region. (b) 2-D surface plot of the polarization profile. Polarization is maximum at the metal/FE interface and minimum at FE/DE interface. (c) Schematic of polarization wave model with an upward and downward domain. (d) Characteristics of y-directional function f(y) used to incorporate gradient in the polarization along the vertical direction.  
		Default parameters: L = 50 nm, $t_{fe}$ = 4 nm, $t_{ox}$ = 1 nm, W = 1 $\mu m$, $V_{gs}$ = $V_{ds}$ = 0 V, $\epsilon_x$ = 22, $\epsilon_y$ = 18 and $\epsilon _{ox}$ = 3.9, $\epsilon _{si}$ = 11.8, $t_1=t_{fe}+t_{ox}$, and $t_2=t_1+t_{si}$.
	 }\label{fig:pol_wave} \vspace{-3mm}
\end{figure*}
\section{Development of Electrostatics model with Multi-Domains}
The electrostatic model of multi-domain FeFET is mainly divided into four subsections. In the first part, the mathematical formulation of the polarization profile is obtained. Subsequently, the polarization wave model is used in the 2-D Poisson's equation to derive the 2-D potential functions in the various regions. Afterward, obtained electrostatic model (electric fields and polarization profile) is used to calculate the net ferroelectric energy. Finally, the domain period and dynamics of domains in the ferroelectric region are captured by minimizing net ferroelectric energy.
\subsection{Polarization wave model}
We assume that polarization of multi-domain in the ferroelectric region is periodic with upward and domain period of $x_a$ and $x_b$, respectively, and transition from upward to downward domain occurs in the finite domain wall width ($d_w$). The periodic texture of domains and finite domain wall width are valid approximations observed in experiments and numerical phase-field simulations \cite{Bratkovsky}-\cite{Luk_yanchuk},\cite{Park}-\cite{Saha_3}. Therefore, the mathematical formulation of the polarization profile is obtained in Fourier series form.  
\begin{align}
	&P(x,y)=\nonumber\\
	&f(y)\left(\frac{a_0}{2}+\frac{2P_{s}}{d_wT_D} \sum_{j}\frac{A_jcos\left ( k_jx \right )+B_jsin\left ( k_jx \right )}{k_j^2}\right) \label{pol_wave_eq}
\end{align}

where, $T_D$ is the total domain period ($x_a+x_b$), $d_w$ is the domain wall width, $P_s$ is the spontaneous polarization and all other Fourier series coefficients are given in the Appendix section.

The y-directional function $f(y)$ is included in (\ref{pol_wave_eq}) to incorporate the vertical directional gradient in domain polarization. 
\begin{align}
	f(y)=\left(\frac{-1}{t_{fe}^{2}}\right)y^2+1
\end{align}

Due to the gradient in polarization along the y-direction, we formulate the function $f(y)$ such that: it is maximum at the FE/metal interface ($y=0$), and it is minimum at the FE/DE interface ($y=t_{fe}$). Therefore, the polarization of a domain is maximum at the metal/FE interface and approaches zero at the FE/DE interface (also observed in phase-field simulations, \cite{Park}, \cite{Saha_1}-\cite{Saha_3}). Fig. \ref{fig:pol_wave} (a) and (b) show polarization's 3-D and surface distribution in the FE region. Fig. \ref{fig:pol_wave}(c) shows the schematic of polarization wave with a domain wall width $d_w$. Fig. \ref{fig:pol_wave}(d) shows the characteristic of $f(y)$.

\subsection{Electrostatic Potential model}
 Fig. \ref{fig:str_1}(a) shows the schematic diagram of the double gate multi-domain FeFET. The 2-D Poisson's equation in the various regions is expressed as.
 \\
 
1. Ferroelectric region ($0 < y <  t_{fe}$)
\begin{align}
\epsilon_x \frac{\partial ^2 \phi_{fe}}{\partial x^2} +\epsilon_y \frac{\partial ^2 \phi_{fe}}{\partial y^2}=-\frac{1}{\epsilon_0}\left(\frac{\partial P(x,y)}{\partial x}+\frac{\partial P(x,y)}{\partial y}\right)
\label{poisson_fe}
\end{align}

2. Insulator region ($t_{fe} < y < t_{fe}+t_{ox} $)
 \begin{align}
 \frac{\partial ^2 \phi_{ox}}{\partial x^2} +\frac{\partial ^2 \phi_{ox}}{\partial y^2}=0\label{poisson_ox}
 \end{align}
\;\;\;\;3. Channel region ($t_{fe}+t_{ox} < y <t_{fe}+t_{ox} +t_{si}  $) 
 \begin{align}
 	\frac{\partial ^2 \phi^{si}}{\partial x^2} +\frac{\partial ^2 \phi^{si}}{\partial y^2}=\frac{qN_a}{\epsilon_{si}}\label{poisson_si}
 \end{align}
This work aims to model the conduction band barrier's oscillations in a FeFET, which are only significant in the weak inversion region. Therefore, we assume the depletion approximation in the channel $ \left(\ref{poisson_si}\right)$.
The elementary task is to capture domain nucleation and dynamics in the FE region.  Domain dynamics incorporate vertical and lateral gradients in polarization waves (\ref{poisson_fe}). The y-directional gradient induces the bound charges ($\rho=-{\partial P}/{\partial y}$) at the FE-DE interface. On the other hand, the lateral gradient in the polarization contributes to domain wall energy density. We will show that various ferroelectric energy components interplay determines a thermodynamically stable domain state in the FE region.

Eq. (3)-(5) are a system of 2-D non-homogeneous partial differential equation (PDE). Green's function approach obtains the analytical solutions of these equations. Earlier, Green's function method is used to obtain the 2-D potential distributions of the various DG-MOSFET structures \cite{Jackson}-\cite{nilesh}. In our recent work, an analytical model of MFIS-NCFET is reported\cite{nilesh}. However, the only mono-domain state is considered in the FE region. Hence, the PDE in FE region reduces to a Laplace equation $\left({\partial P}/{\partial x}={\partial P}/{\partial y}=0\right)$. Therefore, the developed model in \cite{nilesh} cannot capture and study the domain dynamics in the ferroelectric region.

Here, for the first time, we report an analytical and explicit model for multi-domain FeFET with x-y gradients in the polarization (\ref{poisson_fe}).
\begin{align}
\phi ^{fe}(x,y)=\phi_{G_x}(x,y)+\phi_{G_y}(x,y)+\phi^{fe} _{\rightleftharpoons}(x,y)+\phi^{fe}_{\pm }(x,y) \label{pot_fe}
\end{align}

Potential components $\phi_{G_x}(x,y)$, and $\phi_{G_y}(x,y)$ are obtained by following Green's identity.
 
 \begin{align}
 	 G_i=\iint \left ( \frac{\partial P(x,y)}{\epsilon_x\partial x}+\frac{\partial P(x,y)}{\epsilon_y\partial y} \right ){}G(x,y;x',y')dx'dy' \label{identity}
 \end{align}
\begin{figure*}[!t]
	\begin{align} 
		\nonumber
		&\phi_{G_x}(x,y)=\left ( \frac{4P_s}{t_{fe}\epsilon_yd_wT_D } \right )\sum_{n}\frac{sin\left ( k_n^Iy \right )}{sinh\left ( k_n^IL \right )}\left \{ \lambda^n\sum_{j}\left ( \frac{B_j\xi^{j,n}(x)-A_j\zeta^{j,n}(x)}{k_j} \right ) +\Lambda^n\sum_{j}\left ( \frac{A_j\xi^{j,n}(x)+B_j\zeta^{j,n}(x)}{k_j^2} \right )\right \} \\
		&+\left ( \frac{a_0}{t_{fe}\epsilon_y } \right )\sum_{n}\frac{sin\left ( k_n^Iy \right )\Lambda^n}{ \left(k_n^I\right)^2}\left \{ 1-\left ( \frac{sinh\left ( k_n^Ix \right )+sinh\left ( k_n^I\left ( L-x \right ) \right )}{sinh \left ( k_n^IL  \right )} \right ) \right \} \label{phi_Gx} \\
		&\phi_{G_y}(x,y)=\left ( \frac{4P_s}{L\epsilon_xd_wT_D } \right )\sum_{m}\frac{sin\left ( k_mx \right )}{k_m}\left \{ \xi^m(y)\sum_{j} \left( \frac{B_j\lambda^{m}-A_j\Lambda^{m}}{k_j}\right)+ \zeta^m(y)\sum_{j}\left(\frac{A_j\lambda^{m}+B_j\Lambda^{m}}{k_j^2}\right)\right \} \label{phi_Gy}\\ \nonumber \\ \nonumber
		&\text{$\zeta^{j,n}(x)$, and $\xi^{j,n}(x)$ are function of $x$ with two series indices $n$, and $j$.}
		\\
		&\xi^{j,n}(x)=\left ( \frac{1}{k_j^2+\left ( k_n^I \right )^2} \right )\left \{ cos\left ( k_jx \right )sinh\left ( k_n^I L \right )-sinh\left ( k_n^I \left (L-x \right ) \right ) -sinh\left ( k_n^Ix \right ) cos\left ( k_jL \right )\right \}
		\\ 
		&\zeta^{j,n}(x)=\left ( \frac{1}{k_j^2+\left ( k_n^I \right )^2} \right )\left (  sin\left ( k_jx \right )sinh\left ( k_n^I L \right )-sin\left ( k_jL \right )sinh\left ( k_n^I x \right )\right )
		\\ \nonumber \\ \nonumber
		&\text{$\zeta^m(y)$, and $\xi^m(y)$ are function of $y$ with $m$ series indices}
		\\
		&\xi^m(y)= \left \{ \frac{-1}{t_{fe}^2}\left [  \frac{y^2}{k_m}+\frac{2}{k_m^3}\left ( 1-\frac{cosh\left ( k_m\left ( t_{fe}-y \right ) \right )}{cosh\left ( k_mt_{fe} \right )} \right ) -\frac{2t_{fe}}{k_m^2}\left ( \frac{sinh\left ( k_my \right )}{cosh\left ( k_mt_{fe} \right )} \right ) \right ]+\frac{1}{k_m}\left ( 1-\frac{cosh\left ( k_m\left ( t_{fe}-y \right ) \right )}{cosh\left ( k_mt_{fe} \right )} \right ) \right \}
		\\ 
		&\zeta^m(y)=\left \{ \frac{-1}{t_{fe}^2}\left [\frac{y}{k_m}-\frac{sinh\left ( k_my \right )}{k_m^2cosh\left ( k_mt_{fe} \right )} \right ] \right \}
	\end{align}
	\par\noindent\rule{\textwidth}{0.4pt}
\end{figure*}
Gradients of polarization wave ${\partial P(x,y)}{/\partial x}$, and ${\partial P(x,y)}{/\partial y}$ are evaluated by (\ref{pol_wave_eq}). Subsequently, evaluated gradients and Green's functions of the FE region are plugged in (\ref{identity}), which leads to the 2-D potential distributions given by $\left(\ref{phi_Gx}\right)$, and $\left(\ref{phi_Gy}\right)$, respectively. 
The remaining two terms $\phi^{fe} _{\rightleftharpoons}(x,y),\;\text{and}\;\phi^{fe}_{\pm }(x,y)$ originated due to the FE region's left/right ($x=0,\;\text{and}\;x=L$) and top/bottom ($y=0,\;\text{and}\;y=t_{fe}$) boundaries, respectively (given in the Appendix section).

The 2-D potential equations of oxide and channel regions are obtained by using Green's functions of respective regions into the Green's identity\cite{Jackson}.

\begin{align}
	&\phi ^{ox}(x,y)=\phi _{\rightleftharpoons}^{ox}(x,y)+\phi_{\pm }^{ox}(x,y)+\phi^{ox}_{n=0}(x) \\ 
		\nonumber
	&\phi ^{si}(x,y)= E_g/2q+ V_{ds}\frac{x}{L}+ \frac{2}{L}\sum_{m=1}^{\infty }sin\left ( k_{m}x\right )D_{sb}^{m,n}\times \\ 
	&\left(\frac{ cosh\left ( k_{m}\left ( t_{2}-y \right ) \right )+ cosh\left ( k_{m}\left ( t_{1}-y \right ) \right )  }{\epsilon_{si} k_{m}sinh( k_{m}t_{si})}\right)
\end{align}

The mathematical expressions of $\phi ^{ox}(x,y)$ and approach to calculating lateral Fourier series $\left(D_{sf}^{m,n}\text{, and }D_{sb}^{m,n}\right)$ coefficients are given in the Appendix. 
\subsection{Energy dynamics in the Ferroelectric material}
The equilibrium (applied bias = 0 V) and non-equilibrium (applied bias $\neq$ 0 V) configuration of the domains are determined by minimizing net ferroelectric energy leading to a thermodynamically stable state. The net energy density of the ferroelectric material is given as \cite{Lines}-\cite{Luk_yanchuk}.
\begin{align}	
	f_{net} = f_{free} + f_{dep} + f_{grad} \label{fnet}
\end{align}
 where, 
 $f_{free}$ is free energy density and expressed in the Taylor series of ferroelectric polarization \cite{Lines}-\cite{Rabe}.
 \begin{align}
 	f_{free} = f_0+\alpha P^2(x,y)+\beta P^4(x,y)+\gamma P^6(x,y) \label{ffree}
 \end{align}

Depolarizing energy density is evaluated by the local distributions of polarization texture and local electric fields \cite{Park}.
	\begin{align}
		\nonumber
	&f_{dep} = \\ &\frac{1}{2}\left[\epsilon_x \left(E^{fe}_{x}(x,y)\right)^2+\epsilon_y \left(E^{fe}_{y}(x,y)\right)^2\right]+ E^{fe}_y(x,y)P(x,y) \label{fdep}
\end{align}
Contributions of local stray fields and polarization in $f_{dep}$ are incorporated by the first and second terms of (\ref{fdep}), respectively.
\begin{align}
	E^{fe}_{x}(x,y)=-\frac{\partial \phi_{fe}(x,y)}{\partial x}
\end{align}
\begin{align}
	E^{fe}_{y}(x,y)=-\frac{\partial \phi_{fe}(x,y)}{\partial y}
\end{align}
Domain polarization of ferroelectric material exhibits gradients along the spatial $x$ and $y$ directions \cite{Park}-\cite{Saha_3}. These gradients in the polarization are incorporated by the ${\partial P(x,y)}/{\partial x}$ and ${\partial P(x,y)}/{\partial y}$ in (\ref{poisson_fe}). The gradient energy density of the ferroelectric material is expressed as.
 \begin{align}
 	f_{grad} = \frac{1}{2}\left[\frac{g_{11}}{2} \left(\frac{\partial P(x,y)}{\partial x}\right)^2 + \frac{g_{44}}{2}  \left(\frac{\partial P(x,y)}{\partial y}\right)^2 \right] \label{fgrad}
 \end{align}
where, the material-dependent ferroelectric coefficients are taken from \cite{Saha_4}. We consider a stress-free condition in the Landau coefficients, and strain is incorporated in the Taylor series coefficients of (\ref{ffree}). This approximation is validated by the numerical phase-field simulations \cite{Saha_1}.
\subsection{Algorithm to calculate domain period for zero and non-zero applied bias}
Firstly, we evaluate the equilibrium domain period ($x_e$) for the zero applied voltages. At zero bias, upward and downward domain, periods will be identical \cite{Bratkovsky}-\cite{Luk_yanchuk}, \cite{Park}-\cite{Saha_3}. Therefore, the total domain period is calculated as.
\begin{align}
	T_{D}=x_a+x_b+4d_w\; (x_a=x_b=x_e) \Rightarrow T_{D}=2x_e +4d_w \label{TD_eq}
\end{align}
The equilibrium domain period is evaluated by minimizing net ferroelectric energy $\left(F_{net}\right)$.
\begin{align}
	F_{net}=W\int_{0}^{t_{fe}}\int_{0}^{L}\left(f_{free} + f_{dep} + f_{grad}\right)dxdy \label{Fnet}
\end{align}
where, $W$ is the width of device. Typical value of domain wall width is $\approx$ 0.5 nm$-$1 nm \cite{Park}-\cite{Saha_3}. Hence, after considering an appropriate value of $d_w$ the only unknown in (\ref{Fnet}) is $x_e$, which is calculated by the minimization of net ferroelectric energy as.  
\begin{align}
	\frac{\partial F_{net}}{\partial x_e}=0 \;(\text{solve for $x_{e}$}) \label{eq_xe}
\end{align}
For a given set of physical parameters such as: $t_{fe}$, $t_{ox}$, $L$, $t_{si}$, and ferroelectric material parameters ($\alpha$, $\beta$, $\gamma$ etc.) the above equation is an only function of $x_{e}$ which is solved analytically.

The final task is to evaluate the domain period for non-zero applied voltages (non-equilibrium state). Non-zero applied bias triggers the domain wall motion \cite{Park}-\cite{Saha_3}. Hence, the period of the upward and downward domain is no longer remains identical ($x_a\neq x_b$). Depending on the polarity of applied voltage, the expansion/reduction in a domain period is observed. An $\delta x$ variation in an upward domain period leads to the same amount of variation in the downward domain period \cite{Park}-\cite{Saha_3}. Therefore, the total domain period remains the same as in an equilibrium condition period (\ref{TD_eq}).
\begin{align}
	T_{D}=x_e+\delta x+x_e-\delta x+4d_w=2x_e+4d_w \label{TD_noneq}
\end{align}
 However, due to the non-equilibrium condition, $x_a\neq x_b$, and the amount of shift is given by $\delta x$,  which is calculated by the minimization of net ferroelectric energy.
\begin{align}
	\frac{\partial F_{net}}{\partial \left(\delta x\right)}=0 \;(\text{solve for $\delta x$}) \label{noneq_delx}
\end{align}

Domain dynamics is captured by plugging $\delta x$ (For each applied voltages) in (\ref{pot_fe}). 
\section{Results and Discussion}
\begin{figure}[!t]
	\centering 
	\centering 
	\includegraphics[width=0.5\textwidth]{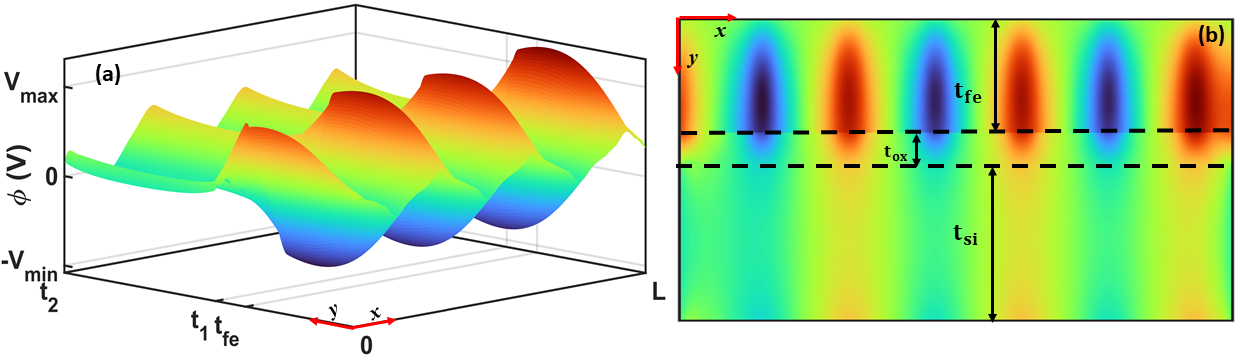}
	\caption{(a) 3-D surface plot of potential band energy. (b) 2-D surface plot of electrostatic potential energy. Source side Fermi level is taken as reference. Parameters: if not mentioned, then default values of parameters are taken.
	}\label{fig:Ec_1}\vspace{-3mm}
\end{figure}

To check the robustness and accuracy of the model, we consider both types of validations: numerical simulations and experimental results. Fig. \ref{fig:str_1}(b) shows validations of the developed model against phase-field simulations \cite{Saha_4}, and experimental results \cite{Krivokapic}.  
Fig. \ref{fig:Ec_1} shows the electrostatic potential distribution (at zero bias) of a FeFET with multi-domains. Due to the adjacent periodic nature of polarization charges in the FE region (see \ref{fig:pol_wave}(a)), the electrostatic of the device also exhibits periodicity. The value of $\Delta \phi$ (difference between maximum and minimum values of the potential) is $\approx$ 500 mV and 40 mV at the metal/FE ($y=0$) interface and at the mid-channel $\left(y=t_{fe}+t_{ox}+t_{si}/2\right)$, respectively. Such significant variations in the channel potential will exponentially alter the thermionic current transport.

\begin{figure}[!b] 
		\centering 
			\includegraphics[width=0.5\textwidth]{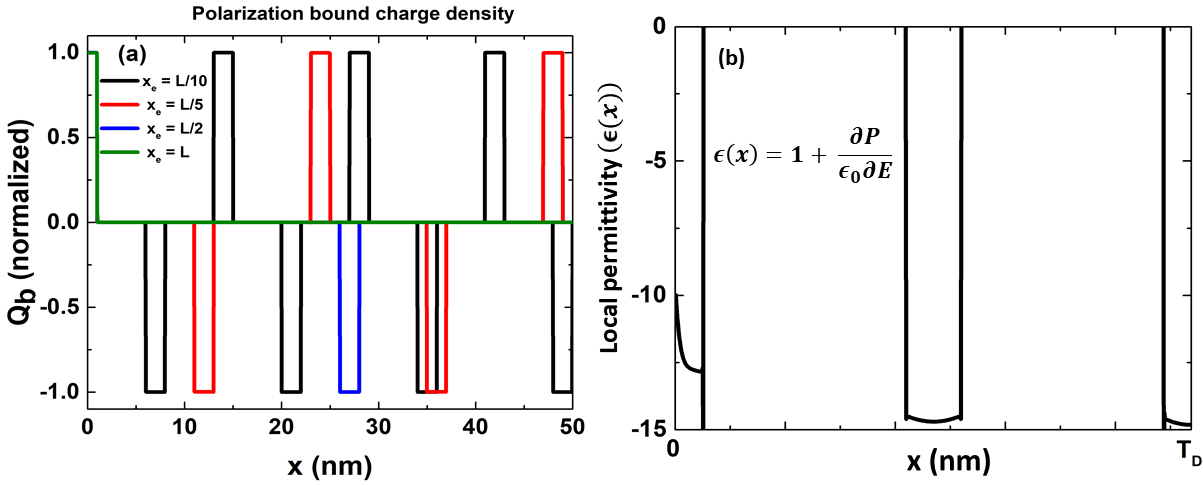}                
	\caption{(a) Bound charge density in the FE region due to multi-domains. (b) Emergence of local negative permittivity caused by domain wall motion.} \label{fig:DW_NC} \vspace{-3mm}
\end{figure}

\subsection{Negative Capacitance via Domain wall motion}
The lateral gradient of polarization $\left(\partial P/ \partial x \right)$ profile in (\ref{poisson_fe}) induces the bound charge density along the lateral direction. Fig. \ref{fig:DW_NC}(a) shows the bound charge density ($Q_b$) distribution in the FE region for various domain periods. Charge density $Q_b$ exhibits the periodic nature with positive and negative values. Fig. \ref{fig:DW_NC}(b) shows the local permittivity of the FE region (for a domain period = $T_D$), which is calculated by the electric displacement vector. The emergence of negative permittivity is observed due to the negative slope of polarization vector w.r.t. electric field vector $\left(\partial P/\partial E< 0\right)$. The signature of negative permittivity is the origin of negative capacitance induced by domain wall motion. Note that, recently, an experimental study also demonstrated the presence of NC by domain dynamics in a ferroelectric capacitor \cite{Yadav}. Therefore, the developed model can capture inherent negative capacitance by domain wall motion.  
 \begin{figure}[!t] 
	\centering 
	\includegraphics[width=0.5\textwidth]{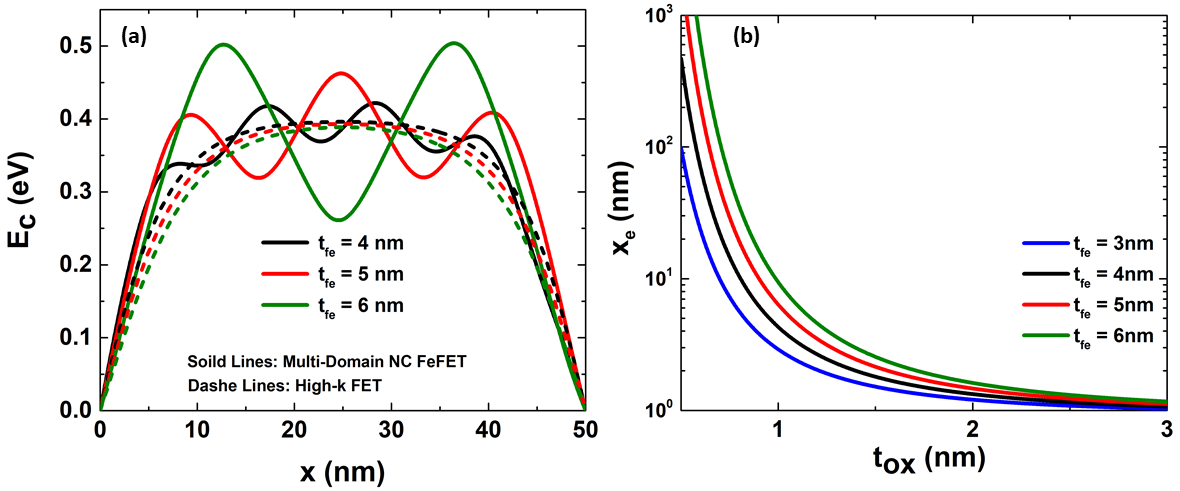}                
	\caption{(a) Conduction band energy plotted at mid-channel. Oscillations in the conduction band are originated due to the multi-domains effect. (b) The transition from mono-domain state to multi-domain state with the variations in physical parameters.} \label{fig:Ec_MD}
\end{figure}

Fig. \ref{fig:Ec_MD}(a) shows the conduction band profile plotted at the middle of the channel for various $t_{fe}$ values.
Due to enhanced fringing fields, the negative capacitance effect shoots up as ferroelectric thickness increases \cite{Phawa}. Therefore, the barrier height in the conduction band rises with an increasing $t_{fe}$.
The number of domains in the FE region is determined by the interactions in various energy components given by (\ref{fnet}). An enhancement in $t_{fe}$ corresponds to a wider dispersion in domain polarization along the vertical direction. Therefore, the slope of $f(y)$ decreases (see Fig. \ref{fig:pol_wave}(d)), which reduces gradient energy density along the y-direction. This reduction in the gradient energy relaxes the compulsion to nucleate denser domain patterns for achieving energy minima. Now, energy can be minimized by the lesser number of domains. Hence, the domain period increases, which reduces the rate of oscillations in the conduction band, as shown in Fig. \ref{fig:Ec_MD}(a). 
On the other hand, as $t_{ox}$ increases, the depolarizing fields increases \cite{Bratkovsky},\cite{Park}-\cite{Saha_3}, \cite{Bratkovsky_2}. Therefore, more number domains nucleates to compensate for the increased depolarizing energy, as shown in Fig. \ref{fig:Ec_MD}(b). Note that similar kinds of observations were also reported in a pioneer work by Bratkovsky et al. \cite{Bratkovsky}, and the phase-field simulations \cite{Park}-\cite{Saha_3}. Therefore, Fig. \ref{fig:Ec_MD}(b) shows a further validation of the developed model and algorithm to capture nucleation of domains in the FE region.  

%Fig. \ref{fig:Ec_tfe}(b) shows the plot of conduction band energy at mid-channel. Increment in $t_{fe}$ enlarges the domain period, therefore, rate of oscillations decreases in the conduction band profile. However, due to increased area of bound charge density in the FE region, magnitude of ripples in the $E_c$ rises as $t_{fe}$ increases. 
\subsection{Electronic transport with multi-domains}	
The drain to source current is calculated by the Quasi-Fermi potential of the channel \cite{nilesh}.
\begin{align}
	I_{ds}=\frac{\mu WkT\left(1-e^{-qV_{ds}/kT}\right)}{\int_{0}^{L}\frac{dx}{\int_{t_1}^{t_1+t_{si}}n_ie^{q\phi_{si}(x,y)/kT}dy}}
\end{align}

\begin{figure}[!t] 
	\centering 
	\includegraphics[width=0.5\textwidth]{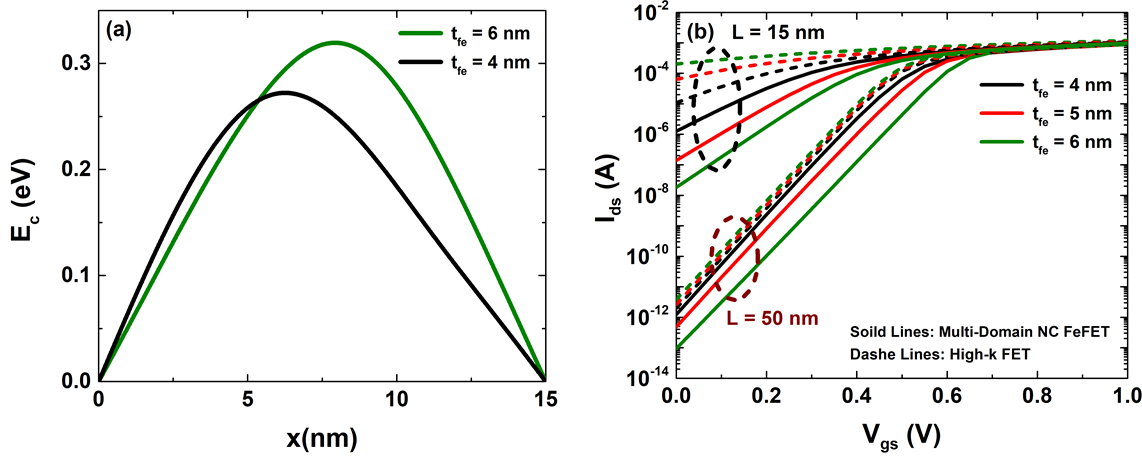}                
	\caption{(a) Mid-channel conduction band plot for the short channel device. A shift in the barrier height occurs due to the nucleation of a domain wall. (b) Drain current characteristics for short and long channel device.} \label{fig:short_MD}\vspace{-3mm}
\end{figure}

Fig. \ref{fig:short_MD}(a) shows a short channel device's conduction band energy plot (L = 15 nm).  The number of domain walls varies with the variations in $t_{fe}$ (see Fig. \ref{fig:Ec_MD}). Hence, the position of local NC shifts with the alteration in $t_{fe}$. Thus, the maximum barrier height of the conduction band exhibits a positional variation with ferroelectric thickness. Shifting in the barrier height signifies the presence of a domain wall at that particular position. Additionally, enhancement in the barrier height with a larger $t_{fe}$ is due to the increased NC effect.
Fig. \ref{fig:short_MD}(b) shows the $I_{ds}-V_{gs}$ characteristics of a FeFET. Solid curves are used for FeFET with multi-domains which includes the negative capacitance effect (due to the domain wall motion), and dashed curves are plotted for the conventional high-k FET where the domain wall does not exist (hence, the NC effect is absent). Enhancement in the $t_{fe}$ raises the conduction band barrier height (see Fig. \ref{fig:Ec_MD}(a) and Fig. \ref{fig:short_MD}(a)), which leads to a reduction in OFF current (NC effect gets stronger with the larger value of $t_{fe}$). On the other hand, conventional high-k FET exhibits the reverse trend with $t_{fe}$, as shown in Fig. \ref{fig:Ec_MD}(a) (dashed curves). The barrier height reduces with an increment in the FE layer thickness. Therefore, the OFF current value rises with an increasing value of $t_{fe}$.  
%Interestingly, the barrier height decreases with the increasing value of $t_{fe}$ which is an opposite trend compared to the long channel device (see Fig. \ref{fig:MD_NC}(b)). The reduction in the barrier height can be explained by analyzing the number of domain walls in the FE region. Short channel devices (L$\leq$ 15 nm) can only accommodate 1-2 domain walls leading to a weaker NC effect (NC effect originates due to domain walls, see Fig. \ref{fig:MD_NC}(a)). However, the device exhibits severe short channel effects (SCEs) at such a small length, dominating the weaker NC effect. Therefore, enhancement in $t_{fe}$ leads to a reduction in the barrier height. However, for the different sets of ferroelectric material parameters, one can achieve only a single (mono) downward domain (negative charge) which may enhance the NC effect over the SCEs, leading to an increment in the barrier height with the FE thickness. 

%On the other hand, the OFF current of the short channel device (L = 15 nm) rises up with an increasing value of $t_{fe}$. This reverse trend with $t_{fe}$ is explained by the dominants SCEs, as shown in Fig. \ref{fig:Id_Vg}(a).

\begin{figure}[!b] 
	\centering \hspace{-3mm} 
	\includegraphics[width=0.5\textwidth]{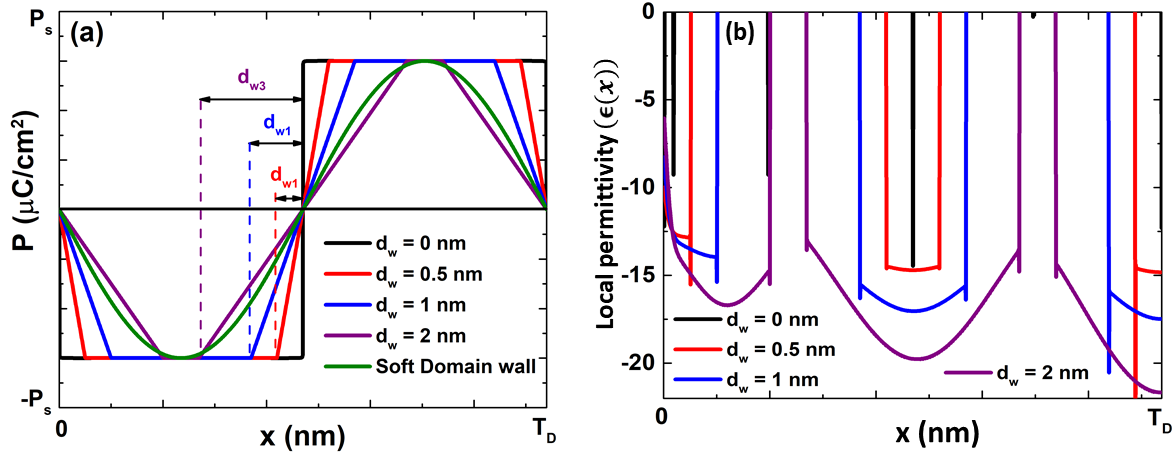}                
	\caption{(a) Transition of domain wall from hard to soft type with increasing value of domain wall width. (b) The magnitude of local permittivity increases with the larger $d_w$.} \label{fig:NC_DW} \vspace{-4mm}
\end{figure}
\subsection{Soft $\rightleftharpoons$ Hard domain wall transition }
\begin{figure}[!t] 
 	\centering \hspace{-3mm} 
 	\includegraphics[width=0.5\textwidth]{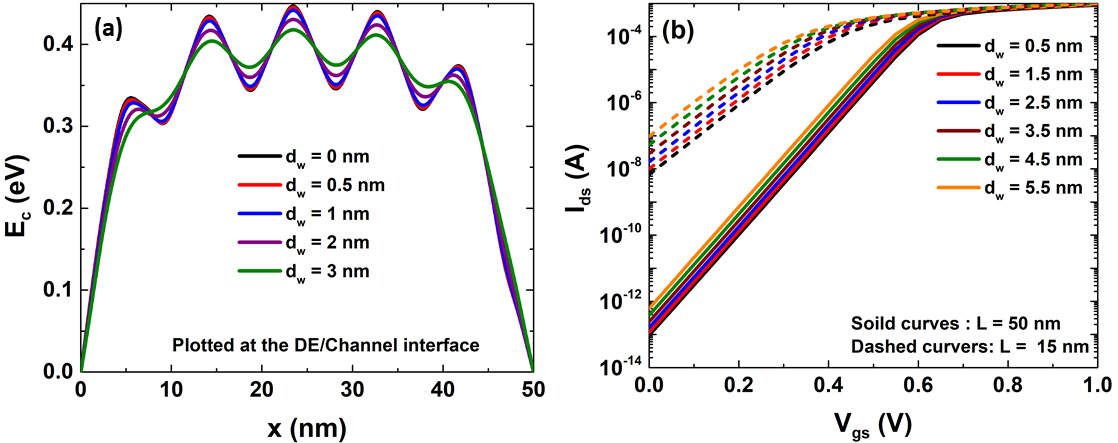}                
 	\caption{(a) Mid-channel conduction band energy for the various value of domain wall widths. Enhancement in $d_w$ leads to a weaker NC effect. (b) Variability in the drain current caused by the domain wall width.} \label{fig:dw_Ec_Id}\vspace{-3mm}
 \end{figure}

The phase-field simulations demonstrated that scaling in the FE layer triggers a domain wall transition from hard to soft type \cite{Saha_1}-\cite{Saha_3}, \cite{Saha_4}. This second order effect is incorporated in the model by transforming the polarization profile in the FE layer. Fig. \ref{fig:NC_DW}(a) shows the polarization profile with the variations in domain wall width ($d_w$). The hard domain wall texture is observed at $d_w\approx 0$. However, a gradual enhancement in $d_w$ leads to a soft domain wall transition. Therefore, transition of the domain wall with $t_{fe}$ can be incorporated by defining a parameter $\alpha_0$.
\begin{align}
	d_w=\frac{\alpha_0}{t_{fe}}d_{w_0}
\end{align}
$\alpha_0$ can be tuned to capture soft/hard domain wall transitions, and $d_{w_0}$ is the starting domain period (without transition). Impact of domain wall transition on negative permittivity is shown in Fig. \ref{fig:NC_DW}(b). The magnitude of local permittivity increases with the larger $d_w$, which can be understood by analyzing the ${\partial P}/{\partial E}$ ratio.
\begin{align}
	\frac{\partial P}{\partial E}=\frac{{\partial P}/{\partial x}}{{\partial E}/{\partial x}}
\end{align}
An increase in $d_w$ reduces ${\partial P}/{\partial x}$ ratio (see Fig. \ref{fig:NC_DW}(a)), however, the ratio ${\partial E}/{\partial x}$ also decreases with the larger $d_w$. 
A larger permittivity value with the $d_w$ states that the electric field vector is more sensitive than the polarization vector for spatial variations $\left(\Delta\left ( {\partial E }/{\partial x} \right )>\Delta\left ( {\partial P }/{\partial x} \right )\right)$. An alternative explanation for increased permittivity with $d_w$ can be understood by observing the conduction band profile, as shown in Fig. \ref{fig:dw_Ec_Id}(a).
As $d_w$ increases, the barrier height decreases. Thus, the NC's strength reduces with the larger $d_w$.
The strength of negative capacitance is inversely proportional to the ferroelectric layer capacitance $\left(\text{NC}\propto 1/\left | C_{FE} \right |\right)$, a larger value of $\left | C_{FE} \right |$ deteriorates the capacitance matching leading to a weaker NC effect \cite{Khan}-\cite{Phawa_5}. 
Since a larger magnitude of local permittivity enhances the $\left | C_{FE} \right |$ value, the barrier height decreases with the increasing $d_w$.
Therefore, the transition towards the soft domain wall increases the sub-threshold current, as observed in Fig. \ref{fig:dw_Ec_Id}(b). Note that variability is higher in the short channel device (L = 15 nm) than the long channel device (L = 50 nm). This may be attributed to the fact that the SCEs dominate at smaller channel lengths, which further reduces the impact of the NC effect, leading to a higher variability.
%Since, the intensity of bound charge density in the FE region is proportional to the polarization gradient $\left(Q_b\propto {\partial P}/{\partial x}\right)$. Therefore, 	
%The slope of the polarization wave increases as the domain wall width increases (see Fig. \ref{fig:NC_DW}(a)). An enhancement in the value of $d_w$  leads to a decrement in the intensity of local bound charges density $\left( {\partial P}/{\partial x}\downarrow\;\Rightarrow\; Q_b\downarrow\right)$. Fig. \ref{fig:NC_DW}(b) shows the conduction band energy plot for the various values of domain wall width. As $d_w$ increases, the strength of the NC effect reduces (NC$\propto Q_b$), which decreases the barrier height of the conduction band. 

%The number of domain walls in the FE region significantly affects the device's electrostatic, as shown in Fig. \ref{fig:NC_DW}. These spatial variations in the electrostatic induce variability in electronic transport. Fig. \ref{fig:dw_Ec_Id}(a) shows the conduction band energy plot for the various values of domain wall width. The increasing value of the $d_w$ leads to a reduction in the NC strength (as explained in Fig. \ref{fig:NC_DW}).
. 

 %	Note that, (\ref{eq_xe}) depends on device geometrical parameters such as $t_{fe}$, and $t_{ox}$. Therefore, alterations in these parameters will lead to variations in $x_e$. Hence, the value of number of domain in the FE region can be controlled by tuning in the thickness of FE/DE layer which corresponds to nucleation/merger of polarizing domains.  
\section{Conclusion}
An analytical physics-based model of multi-domain FeFET is developed by solving 2-D Poisson's equation with the ferroelectric energy dynamics state equation. The process of domain wall nucleation and motion occurs to minimize the net system energy. The dynamics of multi-domains (nucleation of a new domain and domain wall motion) are captured by minimizing net system energy. Ferroelectric energy directly depends on physical parameters such as $t_{fe}$ and $t_{ox}$. Hence, the domain period can be engineered by the alterations in the FE physical parameters. 
The motion of domain wall leads to the origination of local negative permittivity in the FE layer. 
A wider domain wall enhances the magnitude of local permittivity leading to a weaker negative capacitance effect. Furthermore, the transition of domain wall soft$\rightleftharpoons$hard introduces variability in the $I_{ds}-V_{gs}$ characteristics.
%In the long channel device, the NC effect enhances with increasing $t_{fe}$. However, in SCEs, dominant region domain wall induced NC is ineffective, thus, decreases with larger $t_{fe}$.
 
 \section*{Appendix}
$D_{sf}^{m,n}$, and $D_{sb}^{m,n}$ are calculated by the potential continuity condition at FE/DE, and DE/channel interface respectively.
 \begin{align}
 	&D_{sf}^{m,n}=\frac{d_1^m d_\alpha +d_2^md_\beta }{d_4^md_1^m-\left ( d_2^m \right )^2};\;D_{sb}^{m,n}=\frac{d_2^m d_\alpha +d_4^md_\beta }{d_4^md_1^m-\left ( d_2^m \right )^2}
 	 \end{align}
 Note that, the dynamics of multi-domains introduces $d_{MD}^{m,n,j}$ Fourier coefficient in the above equations.
 	 \begin{align}
 	&d_\alpha=d_5^m+d_{MD}^{m,n,j}+d_6^{m,n}-d_8^{m,n}-d_{n=0}^m;\\[0.5em]
 	&d_\beta=d_7^{m,n}+d_{n=0}^m-d_3^m\\[0.5em]
 	&d_{MD}^{m,n,j}=d_{Gx}^{m,n,j}+d_{Gy}^{m,n,j}
 \end{align}
 Fourier series coefficients: $d_i^m\;\&\;d_h^{m,n}$ are evaluated by multiplying sin$\left(k_mx\right)$ in the potential functions and integrating from 0 to L. Where, $i=1,2,3,4,5\;\&\; h= 6,7,8$. \\[0.5em]
$\lambda^n, \Lambda^n, \lambda^m, \text{and} \;\Lambda^m $ are the Fourier series coefficients with $n$ and $m$ indices.
\begin{align}
	&\lambda^n= \frac{2}{t_{fe}^2\left ( k_n^I \right )^3}-\frac{2(-1)^{n+1}}{t_{fe}\left ( k_n^I \right )^2} +\frac{1}{k_n^I}\\[0.5em]
		&\Lambda^n= -\frac{2(-1)^{n+1}}{\left ( k_n^I \right )^2t_{fe}^2}\\[0.5em]
	&\lambda^m=\left ( 1+(-1)^{m+1}cos\left ( k_jL \right ) \right )\left ( \frac{k_m}{k_m^2-k_j^2} \right )
\end{align}
\begin{figure*}[!t] \vspace{-3cm}
	\begin{align}  
	&\Lambda^m=(-1)^{m+1}sin\left ( k_jL \right )\left ( \frac{k_m}{k_m^2-k_j^2} \right )\\[0.5em] 
		&d_{Gx}^{m,n,j}=\left ( \frac{4P_s}{t_{fe}\epsilon_yd_wT_D } \right )\sum_{n}\frac{sin\left ( k_n^It_{fe} \right )}{sinh\left ( k_n^IL \right )}\sum_{j}\left \{ \lambda^n\left ( \frac{B_j\xi^{m,j,n}-A_j\zeta^{m,j,n}}{k_j} \right ) +\Lambda^n\left ( \frac{A_j\xi^{m,j,n}+B_j\zeta^{m,j,n}}{k_j^2} \right )\right \}\\
		&+\left ( \frac{a_0}{t_{fe}\epsilon_y } \right )\sum_{n}\frac{sin\left ( k_n^I t_{fe} \right )\Lambda^n}{ \left(k_n^I\right)^2}\left \{ \frac{(-1)^{m+1}}{k_m}- \frac{k_m}{k_m+\left ( k_n^I \right )^2} \left ( 1+(-1)^{m+1} \right ) \right \} \\[0.5em]
		&d_{Gy}^{m,n,j}=\left ( \frac{2P_s}{\epsilon_xd_wT_D } \right )\sum_{m}\frac{1}{k_m}\left \{ \xi^m(t_{fe})\sum_{j} \left( \frac{B_j\lambda^{m}-A_j\Lambda^{m}}{k_j}\right)+ \zeta^m(t_{fe})\sum_{j}\left(\frac{A_j\lambda^{m}+B_j\Lambda^{m}}{k_j^2}\right)\right \}\\
		&\xi^{m,j,n}=\frac{\left ( 1+(-1)^{m+1}cos\left ( k_jL \right ) \right )}{k_j^2+\left ( k_n^I \right )^2} \left \{ \frac{k_m}{k_m^2-k_j^2}- \frac{k_m}{k_m^2+\left ( k_n^I \right )^2}\right \}\\
		&\zeta^{m,j,n}=\frac{(-1)^{m+1}sin\left ( k_jL \right ) }{k_j^2+\left ( k_n^I \right )^2} \left \{ \frac{k_m}{k_m^2-k_j^2}- \frac{k_m}{k_m^2+\left ( k_n^I \right )^2}\right \}
		\\[0.5em]
		&A_j=cos\left ( k_jd_w \right )-1+cos\left ( k_j\left ( x_1+d_w \right ) \right )-cos\left ( k_j\left ( x_1+3d_w \right ) \right )+cos\left ( k_j\left ( x_1+x_2+4d_w \right ) \right )-+cos\left ( k_j\left ( x_1+x_2+3d_w \right ) \right )\\
		&B_j=sin\left ( k_jd_w \right )+sin\left ( k_j\left ( x_1+d_w \right ) \right )-sin\left ( k_j\left ( x_1+3d_w \right ) \right )+sin\left ( k_j\left ( x_1+x_2+4d_w \right ) \right )-sin\left ( k_j\left ( x_1+x_2+3d_w \right ) \right )\\[0.5em]
		&\phi _{\rightleftharpoons}^{fe}(x,y)= \frac{2}{L}\sum_{m=1}^{\infty }\frac{sin\left ( k_{m}x \right )sinh(k_{m}y) D_{sf}^{m,n}}{\epsilon_{fe} k_{m}cosh( k_{m}t_{fe})}+\frac{2}{L}\sum_{m=1}^{\infty }sin\left ( k_{m}x \right )\left(\frac{\left ( 1-(-1)^m \right ) \left ( V_{gs}-\Delta\phi  \right )cosh\left ( k_{m}\left ( t_{fe}-y \right ) \right ) }{k_{m}cosh( k_{m}t_{fe})}\right)\\[0.5em] 
		&\phi_{\pm }^{fe}(x,y)=\frac{2}{t_{fe}} \sum_{n} \frac{sin\left ( k_{n}^{I}y \right )\left ( A_{1}^{n}sinh\left ( k_{n}^{I}\left ( L-x \right ) \right )+A_{2}^{n}sinh\left ( k_{n}^{I}x \right ) \right ) }{ sinh( k_{n}^{I}L)} \\[0.5em]
		&\phi ^{ox}(x,y)=\phi _{\rightleftharpoons}^{ox}(x,y)+\phi_{\pm }^{ox}(x,y)+\phi^{ox}_{n=0}(x) \\[0.5em]  
		&\phi _{\rightleftharpoons}^{ox}(x,y)=\frac{2}{t_{fe}}\sum_{n=1}^{\infty }\frac{cos\left ( k_{n}^{II}\left ( t_{fe}-y \right ) \right )}{ sinh( k_{n}^{II}L)}\left ( B_{1}^{n}sinh\left ( k_{n}^{II}\left ( L-x \right ) \right )-B_{2}^{n}sinh\left ( k_{n}^{II}x \right ) \right )\\[0.5em]
		&\phi_{\pm }^{ox}(x,y)=\frac{2}{L}\sum_{n=1}^{\infty }\frac{sin\left ( k_{m}x \right )\ }{\epsilon_{ox} k_{m}sinh( k_{m}t_{ox})} \left \{  D_{sf}^{m,n}cosh\left ( k_{m}\left ( t_1-y \right ) \right )- D_{sb}^{m,n}cosh\left ( k_{m}\left ( t_{fe}-y \right ) \right )  \right \} \\[0.5em]
		&\phi^{ox}_{n=0}(x)=\frac{1}{2t_{ox}L}\left ( t_1^2-t_{fe}^2 \right ) \left ( b_{1s}\left ( L-x \right )+xb_{1d} \right )+\frac{1}{2L}\left ( \left ( L-x \right )b_{2s}+xb_{2d} \right )
		\\[0.5em] \nonumber
		&\text{where, $A_{1}^{n}, A_{2}^{n}, B_{1}^{n}$, and $B_{2}^{n}$ are the Fourier coefficients evaluated at the oxide gaps.} \\[0.5em]
		&A_1^n=\frac{\left(V_{gs}-\Delta \phi\right)}{k_n^I}+a_{1s}\frac{sin\left(k_n^I t_{fe}\right)}{\left(k_n^I\right)^2\left({1+\epsilon_{y}/\epsilon_{x}}\right)};\;B_1^n=b_{1s}\frac{\left((-1)^n-1\right)}{\left(k_n^{II}\right)^2}\\
		&a_{1s}=\frac{E_g/2q-V_{gs}+\Delta \phi +P(0,t_{fe}/2)t_{ox}/\epsilon_{ox}}{t_{fe}+t_{ox}\left ( \epsilon_{y}/\epsilon_{x} \right )}; \;b_{1s}=a_{1s}\frac{\epsilon_{y}}{\epsilon_{x}}-\frac{P_s}{\epsilon_{ox}};\;b_{2s}=\frac{E_g}{2q}-b_{1s}t_1;\; a_0= \frac{2P_s}{T_D}\left(x_a-x_b\right)\\[0.5em] \nonumber
		&\text{$b_{1d},$ $b_{2d}$, $A_{2}^{n}, \text{and} B_{2}^{n}$ are calculated by the same manner by replacing $E_g/2q$ with $E_g/2q+V_{ds}$. }\\ \nonumber
		&k_m=m\pi/L;\;k_{n}^{I}=\frac{\left(2n-1\right)\pi}{t_{fe}};\;k_{n}^{II}=\frac{n\pi}{t_{ox}};\;k_j=\frac{2\pi j}{T_D};\;t_1=t_{fe}+t_{ox};\;t_2=t_1+t_{si}.\\ \nonumber
		&\Delta \phi = \text{metal-work function, default value = 4.4 eV.}\\ \nonumber
		&\text{Green's function for the various region are derived in our earlier work\cite{nilesh}.}
	\end{align}
	\par\noindent\rule{\textwidth}{0.4pt}
\end{figure*}

\end{document}